\pdfoutput=1
\documentclass[amsmath,amssymb,aps,twocolumn,superscriptaddress]{revtex4-2}
\usepackage{amsmath}
\usepackage{amssymb}
\usepackage{bm}
\usepackage{graphicx}
\usepackage{epsf}
\usepackage{xcolor}
\usepackage{color}
\usepackage{lipsum}
\usepackage{graphicx}
\usepackage[version=4]{mhchem}
\usepackage[bookmarks=false]{hyperref}
\usepackage{siunitx}
\usepackage[english]{babel}
\usepackage[T1]{fontenc}
\usepackage{float}
\usepackage{dcolumn}

\makeatletter
\def\maketitle{
\@author@finish
\title@column\titleblock@produce
\suppressfloats[t]}
\makeatother
\newcommand{\beginsupplement}{
    \setcounter{table}{0}
    \renewcommand{\thetable}{S\arabic{table}}
    \setcounter{figure}{0}
    \renewcommand{\thefigure}{S\arabic{figure}}
    \setcounter{equation}{0}
    \setcounter{section}{0}
    \renewcommand{\theequation}{S\arabic{equation}}
}

\begin{document}
\title{Light-Induced Electron Pairing in a Bilayer Structure}

\author{Qiaochu Wan}

\affiliation{Department of Physics, University of Pittsburgh, 3941 O’Hara Street, Pittsburgh, Pennsylvania 15218, USA}

\author{Daniel Vaz}
\affiliation{Department of Physics, University of Pittsburgh, 3941 O’Hara Street, Pittsburgh, Pennsylvania 15218, USA}

\author{Li Xiang}
\affiliation{National High Magnetic Field Lab, Florida State University, 1800  E. Paul Dirac Dr, Tallahassee, 32310-3706, FL, USA}

\author{Anshul Ramavath}
\affiliation{Department of Physics, University of Pittsburgh, 3941 O’Hara Street, Pittsburgh, Pennsylvania 15218, USA}

\author{Brandon Vargo}
\affiliation{Department of Physics, University of Pittsburgh, 3941 O’Hara Street, Pittsburgh, Pennsylvania 15218, USA}

\author{Juntong Ye}
\affiliation{Department of Physics, University of Pittsburgh, 3941 O’Hara Street, Pittsburgh, Pennsylvania 15218, USA}

\author{Jonathan Beaumariage}
\affiliation{Department of Physics, University of Pittsburgh, 3941 O’Hara Street, Pittsburgh, Pennsylvania 15218, USA}

\author{Kenji Watanabe}
\affiliation{Research Center for Electronic and Optical Materials, National Institute for
Materials Science, 1-1 Namiki , Tsukuba, 3050044, Ibaraki, Japan}

\author{Takashi Taniguchi}
\affiliation{Research Center for Materials Nanoarchitectonics (MANA), National
Institute for Materials Science,1-1 Namiki , Tsukuba, 3050044, Ibaraki,
Japan.}

\author{Zheng Sun}
\affiliation{State Key Laboratory of Precision Spectroscopy, East China Normal
University,500 Dongchuan Rd, Shanghai, 200241, Shanghai, China}

\author{Dmitry Smirnov}
\affiliation{National High Magnetic Field Lab, Florida State University, 1800  E. Paul Dirac Dr, Tallahassee, 32310-3706, FL, USA}
\author{Nathan Youngblood}
\affiliation{Department of Electrical and Computer Engineering, University of
Pittsburgh, 3941 O’Hara Street, Pittsburgh, 15212, PA, USA.}

\author{Igor V. Bondarev}
\affiliation{Department of Mathematics and Physics, North Carolina Central University,
1801 Fayetteville Str, Durham, 27707, NC, USA.}

\author{David W. Snoke}
\thanks{Address correspondence to: qiw74@pitt.edu, snoke@pitt.edu}
\affiliation{Department of Physics, University of Pittsburgh, 3941 O’Hara Street, Pittsburgh, Pennsylvania 15218, USA}

\date{\today}

\begin{abstract}
We demonstrate the existence of doubly charged exciton states in strongly screened bilayers of transition metal dichalcogenide (TMD) layers. These complexes are important because they are preformed electron pairs that can, in principle, undergo Bose-Einstein condensation (BEC), in which case they would also form a new type of superconductor, consisting of stable bosons with net charge. Our measurements include 1) continuous control of the doping density with both positive and negative carriers, showing the expected population dependencies on the free carrier density, and 2) measurement of the dependence on magnetic field, showing that this new bound state is a spin triplet. These results imply that it is promising to look for superconductivity in this system.

\end{abstract}

\maketitle

\section{INTRODUCTION}
\begin{center}
\begin{figure*}[!htbp]
\centering    

\includegraphics[width=0.9\textwidth]{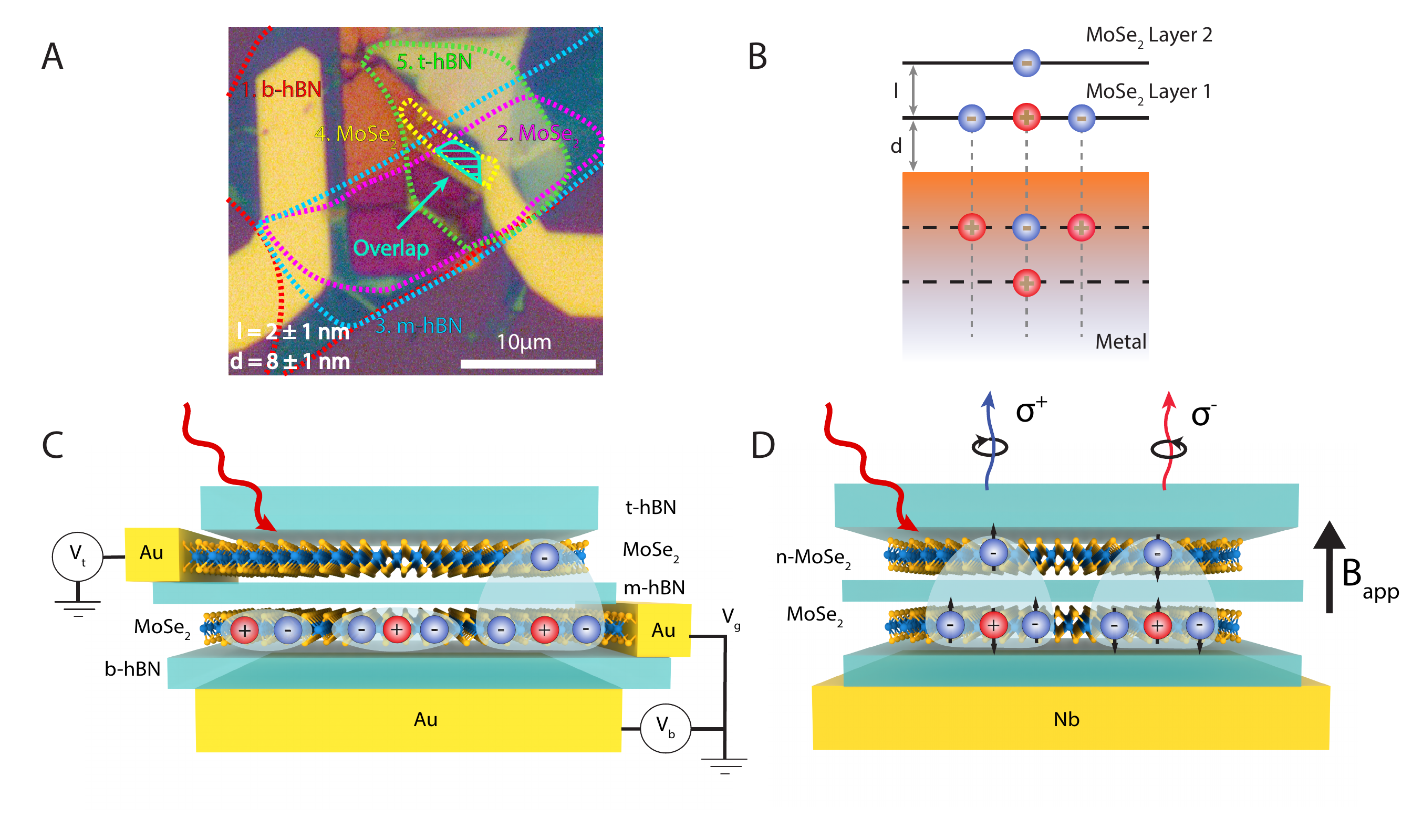}

\caption{\textbf{Sample design:} \textbf{(A)} Microscope image of the quaternion sample. \textbf{(B)} Illustration of the quaternion four-fermion state, with image charge in the metal layer. \textbf{(C)} The design of the structure, showing the circuit connections, with the middle layer defined as the ground. \textbf{(D)} Illustration of the spin states leading to the different circular polarizations of emitted light.}\label{Fig1}
\end{figure*}
\end{center}

Can electrons be paired into bosonic states by any mechanism other than Cooper pairing? While several mechanisms have been proposed, none has been clearly demonstrated. However, a novel proposal by by V.I.~Yudson \cite{yudson1996charged} 
showed how the use of light pumping could produce metastable electron pairing by coupling two electrons (or two holes) to an optically-generated exciton. While one might expect that the excess of charge would make this complex unstable, it can be made stable, first by using TMD layers, which have large (ca. $\sim$200 meV) exciton binding energy \cite{chernikov2014exciton}, encapsulated by insulating hexagonal boron nitride (hBN) layers that prevent tunneling current \cite{britnell2012electron}; and second, by placing a metal layer near the TMD structure, which acts to screen out much of the repulsion between like charges, as illustrated in Figure~\ref{Fig1}B. Because they carry two net charges and are also bosonic, a Bose-Einstein condensate (BEC) of these would be a superfluid, and therefore also a Schafroth superconductor \cite{schafroth1954theory,lozovik1976new}. These complexes can be called tetrons \cite{hadizadeh2013effective} or quaternions \cite{moskalenko2000bose}; here we use the latter term. 

A metastable state of four fermions in a 2D structure is no longer controversial, in general. In the past decade, numerous reports have confirmed the existence of higher-order excitonic complexes in transition metal dichalcogenide (TMD) monolayer and bilayer structures. These include both intralayer and interlayer excitons \cite{jiang2021interlayer}, trions (excitons bound to one extra charge) \cite{plechinger2015identification}, biexcitons (two neutral excitons bound together) \cite{li2018revealing}, hexcitons (three excitons bound together) \cite{choi2023emergence} and charged biexcitons (biexcitons bound to one extra charge) \cite{li2018revealing}. The existence of a state with net charge of $2e$, however, has been less well established.  Previous work \cite{sun2021charged} showed evidence of the four-particle bound state proposed by Yudson, but was limited only to studies in intrinsically doped samples.

Here we report definitive confirmation of this excitonic electron pairing. In one set of measurements, the density of the background free charge carriers was continuously varied via gating, which allows the comparison of the relative intensities of the quaternion, trion, and exciton lines. Since the quaternion must add an extra charge to a trion to form, its intensity should increase relative to the trion as the total background charge density increases, just as the trion intensity increases relative to the exciton \cite{ross2013electrical}. A second set of measurements in magnetic field shows that the quaternion species is a spin triplet, with three states $m_J = 0, \pm 1$, as expected by considering the spin structure of the underlying electron and hole states, as opposed to the trion and exciton states, which have just two states that emit light.  


\onecolumngrid
\twocolumngrid

\section{Experiment method}



Figure \ref{Fig1}A shows a photograph and
Figure \ref{Fig1}B shows the design of the structure used for the electrical gating experiments. Figure \ref{Fig1}C shows the equivalent circuit. The bottom, metallic (Au) layer has two roles; first, it acts as a back gate for the doping of MoSe$_2$ layers, and second, it provides screening that leads to image charge, as shown in Figure \ref{Fig1}(B). This dual gate structure is aimed to dope the two layers of MoSe$_2$ separately. This structure is commonly used in many interlayer studies \cite{zhou2021bilayer,sung2020broken,wang2021moire}. 
We applied the PDMS dry transfer technique to create the stack \cite{castellanos2014deterministic}; the stacks are built on the gold gate with an hBN layer of approximately 2 nm between the TMD layers, a spacer hBN layer of approximately 8 nm between the metal and the ﬁrst TMD monolayer, and a thick capping hBN layer. The drain and top gate (source) gates are 130 nm gold deposited by EBPG after the transfer.


As shown in previous theory \cite{sun2021charged}, this image charge is crucial for canceling out much of the like-charge repulsion and making the doubly-charged complex stable. Another crucial aspect is to have thin hexagonal boron nitride (hBN) layers encapsulating all of the TMD monolayers; these insulating layers should be thin enough to allow Coulomb interaction between layers while thick enough to prevent tunneling current. The thickness of the hBN layers in all cases was known within an uncertainty of $\pm 1$~nm.

The vertical gating allows us to vary the free charge density in the bilayer region continuously. When the background charge is negative, an exciton can be bound to two free electrons, while if the background is positive, an exciton can bind to two free holes. The intrinsic doping for the MoSe$_2$ is n-type \cite{kang2013band}; therefore we are able to observe the negatively-charged quaternion PL peak with no applied gate voltage.

There was an offset in the open-circuit voltage caused by the intrinsic doping so that the structures had an asymmetric $I-V$ current characteristic (see Supplemental Material). Electrical contact to the bottom MoSe$_2$ was grounded, acting as a drain, and the top gate contacting the top MoSe$_2$ layer could have an applied voltages like a source. This source-drain configuration allowed us to vary the charge in the bilayer without obscuring the imaging of the sample, or changing the asymmetric screening effect of the metallic back gate. 

For the magnetic field experiments, a similar structure was used, also with MoSe$_2$ but without the electrical contacts. The top layer of MoSe$_2$ was an $n$-type monolayer doped with Rhenium. In all experiments, the structures were cooled to approximately 10~K, and optically pumped non-resonantly with a wavelength short compared to the PL emission.
\onecolumngrid
\begin{center}
\begin{figure}[!htp]
\centering\includegraphics[width=0.7\textwidth]{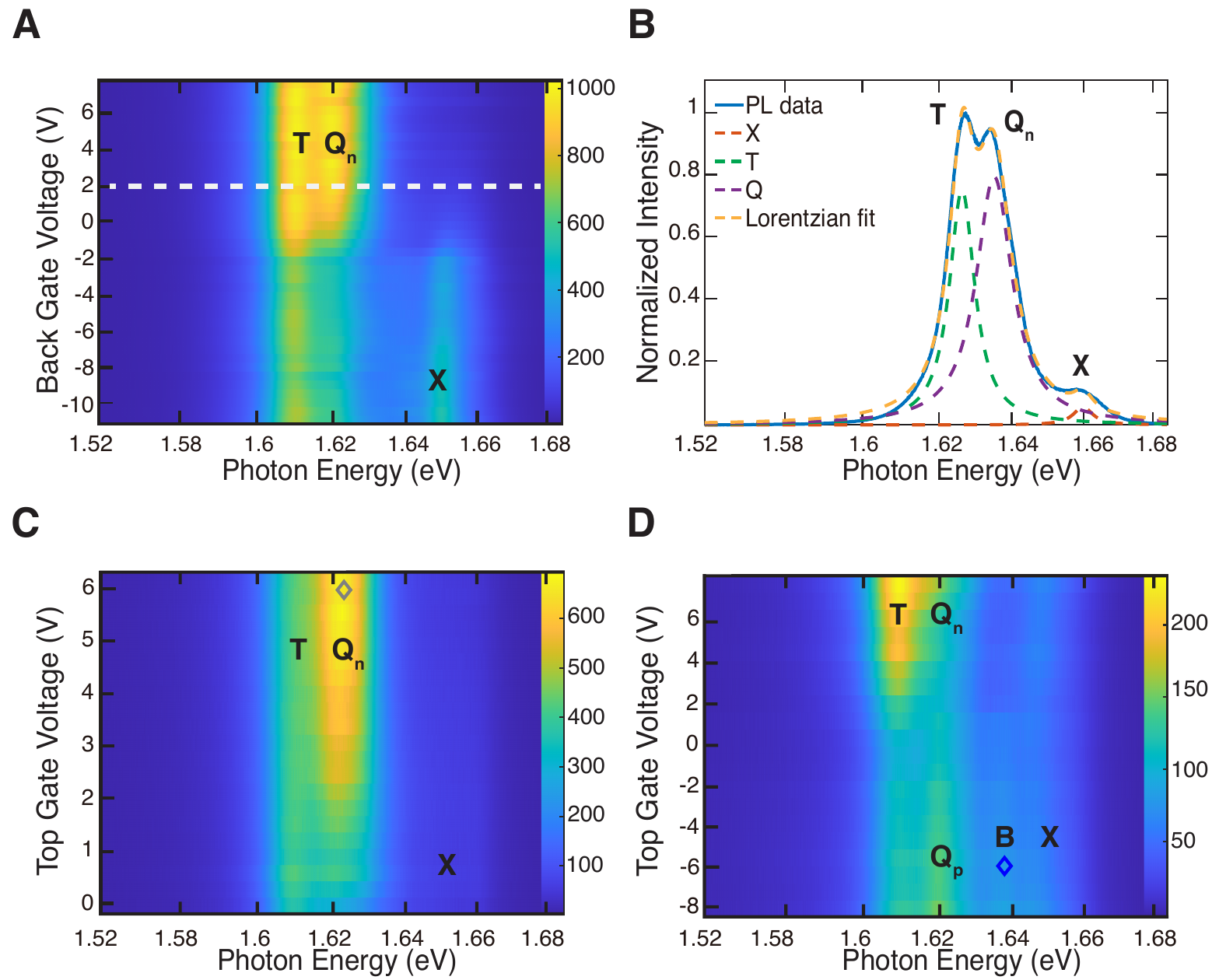}
\caption{\textbf{PL measurements with variable doping:} \textbf{(A)} PL intensity with back gate sweep; the top gate was set to $V=0$. The doubly charged exciton peak, labeled $Q_n$, appears between the trion and exciton lines. 
\textbf{(B)} PL spectrum for a back gate voltage of +2V and top gate voltage of 0V.  This PL is a slice from the white dashed line shown in (A). 
\textbf{(C)} PL with top gate sweep and $V_B = +4$~V. 
\textbf{(D)} PL with top gate sweep and $V_B = - 5$~V. In this case, the trion and quaternion lines reach a minimum, and then increase as positive versions of both appear when the bilayer region is hole-doped. The bath temperature in all of these cases was 5.5 K. (The diamond symbols in (C) and (D) give the conditions for the intensity-dependence measurements of Figures~\ref{Fig3}C and \ref{Fig3}D, respectively.)
}\label{Fig2}
\end{figure}
\end{center}
\twocolumngrid

\section{PL MEASUREMENTS WITH VARIABLE DOPING}
When the structure described above is excited with a laser, an extra PL peak appears between trion and exciton, corresponding to the doubly charged exciton (quaternion) state described above, as seen in earlier work \cite{sun2021charged}. PL data for the structure of Figure~\ref{Fig1} is shown in Figure \ref{Fig2}. 
We applied voltages on both gates of our sample to tune the doping level on the different MoSe$_2$ layers.
The quaternion intensity and trion intensities change dramatically with doping level. We measured the PL with 532~nm green laser pump. For Figures \ref{Fig2}A, we only changed the back gate voltage. For Figures \ref{Fig2}C and Figure \ref{Fig2}D, we fixed the back gate voltage and swept the top gate voltage.

As seen in these figures,  the relative intensities of the quaternions, trions and excitons varied significantly with varying doping levels. Fig.~\ref{Fig2}A illustrates how the PL spectrum varies as the back gate voltage is varied while the top gate is kept grounded. As the overall negative doping density in the bilayer structure is increased, the intensities of trion and quaternion and lines increase, indicating that, like the trion, the quaternion is a charged complex. 
 We note also that the spectral position of the quaternion line does not shift with voltage.
This behavior clearly distinguishes the quaternion from an interlayer exciton, which has an energy position that strongly depends on the electric field between the two layers \cite{szymanska2003excitonic,zhou2021bilayer}. Prior work in a very similar sample has shown that the interlayer exciton line appears at much lower energy  \cite{zhou2021bilayer}.

\begin{figure*}[!htp]
\centering
\centering\includegraphics[width=0.7\textwidth]{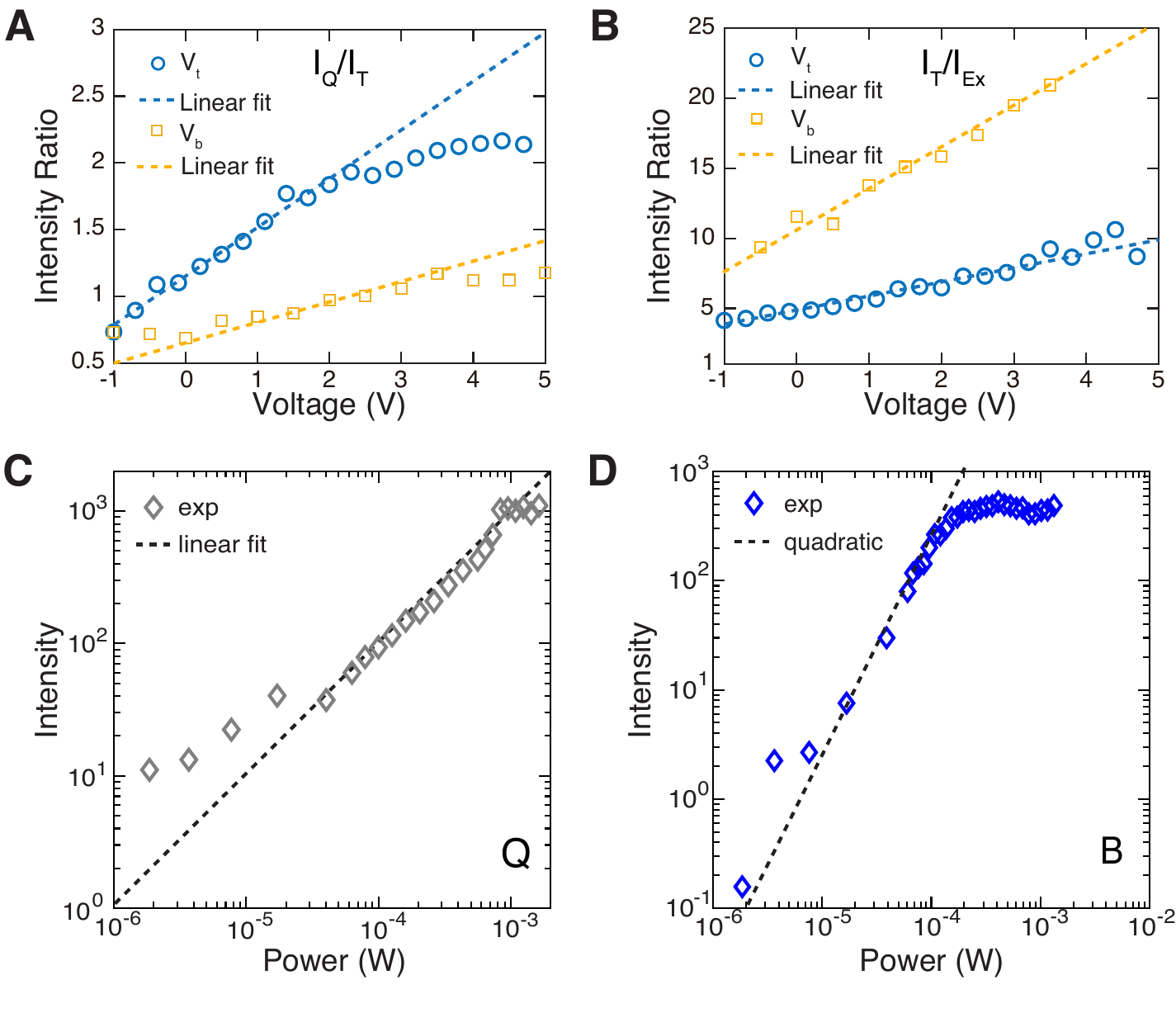}
\caption{\textbf{Voltage and power dependent behavior:} \textbf{(A)} Ratio of the integrated intensity of the quaternion line to that of the trion line as the gate voltages are swept, at constant pump intensity. 
\textbf{(B)} Ratio of the integrated intensity of the trion line to that of the exciton line, for the same voltage sweeps as in \textbf{A}.
\textbf{(C)} Quaternion intensity as a function of pump power, for $V_t$ = 6$~V$ and $V_b$ = 4$~V$ (shown as the gray diamond symbol in Figure 2\textbf{C}).  \textbf{(D)} Biexciton intensity as a function of pump power, for  $V_t = -6$~V and $V_b = -5$~V (shown as the blue diamond symbol in Figure 2\textbf{D}.)  }
\label{Fig3}
\end{figure*}

 Figure~\ref{Fig3}A shows the ratio of the quaternion intensity to the trion intensity as the gate voltages are varied, and Figure~\ref{Fig3}B shows the ratio of the trion intensity to the exciton intensity for the same sweeps.
 The trion line increases relative to the exciton as the background carrier density increases, and the quaternion line increases relative to the trion line under the same conditions, indicating that it has more charge than the trion.  At high gate voltage, the curves in Figures~\ref{Fig3}A and \ref{Fig3}B saturate. This may be due to fermionic state-filling effects.

As seen in Figure~\ref{Fig2}D, we can switch between negative and positive quaternions in a single sweep as the sign of the background carriers is changed; between these two limits, there is a range around top gate voltage of -2V with almost zero quaternion density; in this voltage range there are very few background carriers to pick up. We also see in this figure a weak biexciton line \cite{hao2017neutral}, which is distinguishable from the quaternion line. As seen in Figure~\ref{Fig3}D, the intensity of this biexciton line increases quadratically with pump power, as expected since biexciton creation is a two-photon process.
By contrast, as seen in Figure~\ref{Fig3}C, the Q line intensity is linear with pump power, until high density, where it saturates, presumably due to nonradiative Auger-like processes. 
This shows clearly that the Q line is not a biexciton state. 

\section{MAGNETO-OPTICAL MEASUREMENTS}
\begin{figure*}[!htp]
\centering
\includegraphics[width=0.7\textwidth]{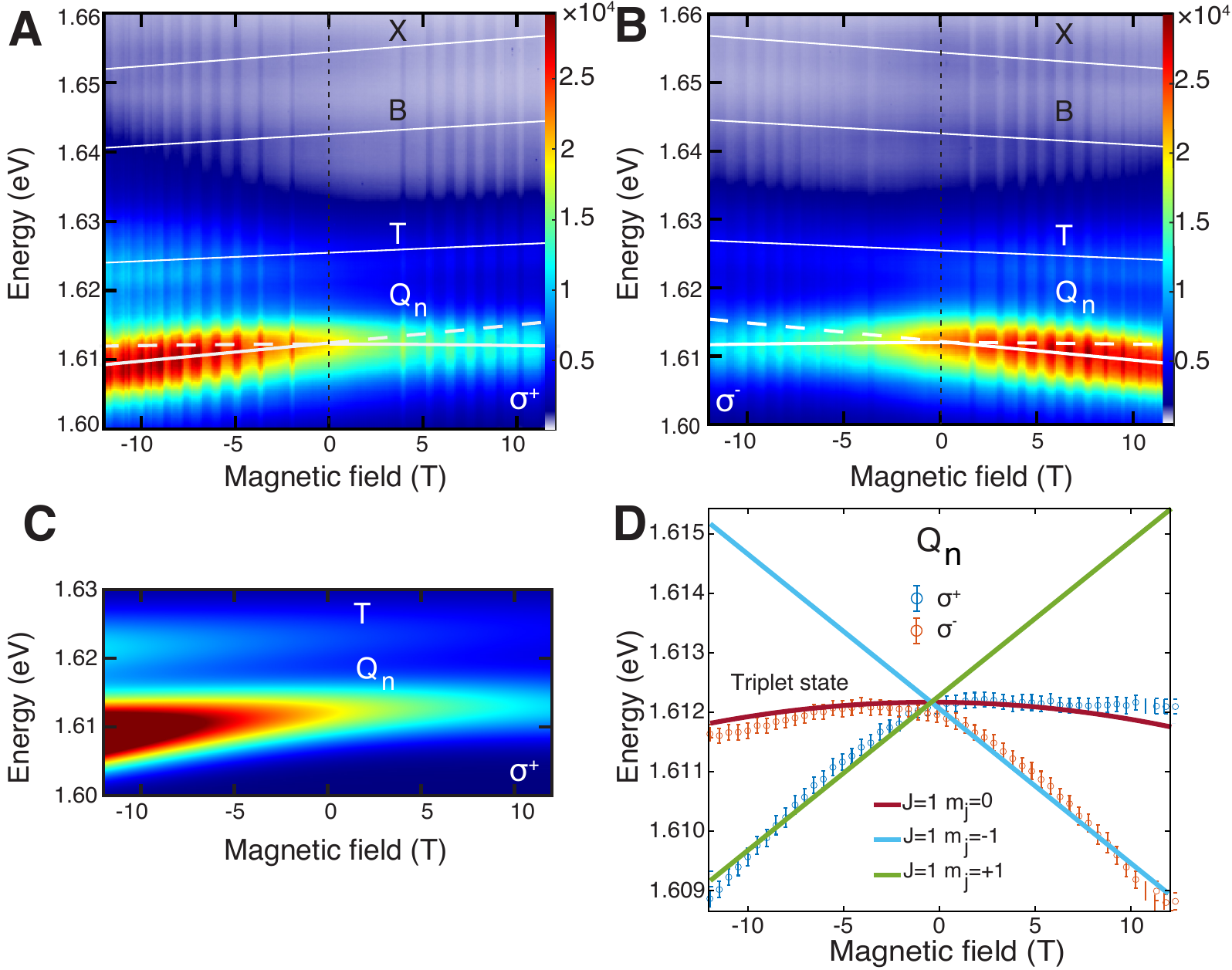}
\caption{\textbf{Magneto-PL measurements:} \textbf{(A)} PL spectrum as a function of magnetic field for a right-circular polarized emission. (The vertical striations correspond to laser intensity fluctuations.) X: exciton line; T: trion line;  B: biexciton line, Q: quaternion line. The white lines indicate linear shifts of the lines with magnetic field. \textbf{(B)} PL spectrum for left-circular polarized PL for the same conditions, with the slopes of the line shifts, but time reversed $B \rightarrow -B$. \textbf{(C)} Simulated data for the right-circular emission from the quaternion triplet, using the linear energy shifts shown in \textbf{(A)} above, weighted by the Maxwellian occupation factor $e^{-E/k_BT}$, with $T = 29$~K, and line width of 4.6 meV.
\textbf{(D)} Composite fit of the full theory (discussed in the Supplemental Material) to the peak energies of the Q line extracted from the right- and left-handed data of \textbf{(A)} and \textbf{(B)}. 
}\label{Fig4}
\end{figure*}

In addition to the above doping-dependent measurements, we have measured the magnetic-field dependence of the photoluminescence (PL) of these complexes. As described above, a $n$-type sample with Rhenium doping was used, without electrical contacts. The magnetic field was applied out-of-plane and swept from -12 to +12 Tesla, and the PL was measured for both circular-polarized components. (The pump polarization was right-circular, but since it was far above the exciton energy, the polarization of the injected carriers was almost certainly lost to scattering processes.) Figure~\ref{Fig4}A shows the spectrum in the range of the intralayer PL emission, as a function of magnetic field for the right-handed polarization (co-circular with the pump) and Figure~\ref{Fig4}B shows the data under the same conditions for left-handed circular polarization.



Several effects are immediately observable in these figures. First, the exciton, biexciton, and trion all have linear shifts with magnetic field as previously observed \cite{macneill2015breaking,aivazian2015magnetic,li2014valley}.
Second, the quaternion line has PL consistent with a spin triplet with all three states having allowed photon emission. In this sample, the quaternion line appears below the trion line, consistent with earlier observations \cite{sun2021charged} that the Q line energy is a strong function of the thickness of the bottom hBN layer, which affects the screening from the metal of the back gate. As seen in Figures~\ref{Fig4}A and \ref{Fig4}B, the quaternion line does not have a simple linear shift. Instead, it can be fit to three lines corresponding to $m_J = \pm 1$ lines that only appear on one side in each of the polarization-resolved images, and an $m_j=0$ state that emits equally with both circular polarizations, and shows no energy shift with magnetic field. This is
consistent with the selection rule analysis given in the Supplemental Material. 

Figure~\ref{Fig4}C shows simulated data which is simply the sum of the right-circular $m_J = 1$ emission plus 50\% of the non-shifting $m_J =0$ emission, with each line weighted by a Maxwellian thermal occupation factor, which makes the lower energy range much brighter. The right-circular emission does not appear above the $m_J=0$ line because it it is thermally suppressed by the Maxwellian factor. For the same reason, we also do not see evidence of a singlet $m_J = 0$ quaterion line at higher energy.

These results, and the identification of the Q line as a quaternion, 
are consistent with a straightforward theory based on the configuration-space method of calculating the binding energies of the excitonic complexes. The same method gives the experimentally confirmed binding energies for both the interlayer trions and for biexcitons in TMD materials \cite{bondarev2018complexes}, and has also been able to explain the evidence for a positive/negative trion binding energy difference \cite{bondarev2021crystal}. In prior work \cite{sun2021charged} we used this method to calculate the quaternion binding energy with the additional inclusion of the image charges in the metal layer.

The details of the model are given in the Supplemental Material, but the essential results needed to understand the quaternion line shifts are straightforward. First, the ground state of the trion has a hole with two electrons in the same spin state. (We consider here the case of an $n$-doped system, corresponding to the sample used in the magnetic field measurements.) This makes the trion an overall spin-1/2 complex. Next, this intralayer trion is a stable ``core'' which is attached to an electron in the other monolayer, to form the quaternion complex. The total quaternion complex therefore is comprised of two spin-1/2 entities, forming triplet and singlet states. 

The ground state of this quaternion complex is the triplet state, consistent with the well-known empirical Hund’s rule of atomic physics~\cite{landau2013quantum}. This rule states that an atomic state with the greatest possible value of the total spin S (for the given electron configuration) and the greatest possible value of angular moment L (for this S) is the one that has the lowest eigenenergy. As an example, in a system of two electrons their total spin can be either 0 or 1. Spin 1 corresponds to a symmetric spin wave function and an antisymmetric coordinate wave function. Since the latter has a node, the probability of finding the two electrons close together is smaller in their triplet state than in the singlet one. This makes their electrostatic repulsion smaller and the eigenenergy lower compared to that of the singlet configuration. Similarly, for a system of several electrons, the most antisymmetric coordinate wave function corresponds to the greatest spin and the lowest eigenenergy. This applies directly to our quaternion system as well, in which in the triplet configuration the electron of the top layer prefers to stay as far as possible from the trion electrons in the bottom layer, so that their repulsion energy is less. 

From the above analysis, the primary shifts of the states will be linear Zeeman-type terms, but as shown in Figure~\ref{Fig4}D, the full theory gives a weak magnetic field dependence of the $m_J = 0$ quaternion state. While this dependence is weak, the fits to the line positions are consistent with the full theory.

\section{Conclusions}

These experimental results present a highly convincing case for the existence of doubly charged excitons, a.k.a. quaternions, in these specially designed TMD bilayer structures. The doping density and magnetic field experiments give unique signatures for this state which agree well with theory.

These are therefore preformed electron pairs that can undergo Bose-Einstein condensation at high density or low enough temperature, and such a BEC state would also be a superconductor. It is unclear whether this state should emit coherent light, as is the case for excitonic and polaritonic condensates. In the case of an excitonic or exciton-polariton condensate, the center-of-mass wave function maps directly to the momentum of an emitted photon. In the case of quaternions, the photon emission leaves behind two electrons and therefore does not map directly to the center of mass of the whole complex.
Although we have increased the excitation density of our pump laser, we have not seen evidence for coherent light emission, for example, in spectral line narrowing. 

Condensation and the consequent superconductivity will be favored at low temperature. In these experiments the effective temperature of the quaternion gas tends to get hotter due to the excess energy put into the system by the non-resonant pump, which in turn raises the critical density needed for Bose condensation. At high density, nonradiative collisional Auger recombination may also become important; there is evidence for this in Figures~\ref{Fig3}C and \ref{Fig3}D.
Both of these effects could be reduced by better cooling the sample; for example, in an immersion flow cryostat as opposed to the cold-finder cryostats used in these experiments. Also, it may be beneficial to laterally confine the carriers in a trap, as done, e.g., for polaritons in Ref.~\cite{balili2007bose}. 

Technically, true BEC is not possible in one and two dimensions, but it can often be the case that the coherence length of a two-dimensional Bose system is large compared to the system size. In general, in any finite system in which the size of the system is small compared to the coherence length, the system can undergo a transition indistinguishable from BEC. 
Another fascinating possibility is that since these complexes have long-range Coulomb repulsion, they could form a bosonic Wigner crystal at low temperature, and in principle, could even become a supersolid. The process of Wigner crystallization is controlled by the ratio of the Coulomb repulsion energy over the average kinetic energy of an ensemble of charged particles~\cite{Platzman-Fukyuama}. Due to their double charge and triple mass as compared to electrons, this ratio is at least 10 times greater for quaternions suggesting much higher quaternion Wigner crystallization temperature than that of the order of 10 K recently observed for quasi-2D electrons in TMD nanostructures~\cite{Smolenski-Immamoglu,Zhou-Demler-Immamoglu}.
 
 The establishment of quaternion complexes in bilayer systems with metal screening layers opens up a promising field of research, with the real possibility of finding a new, non-BCS type of superconductivity.


\section{Acknowledgments}

We thank Zhehao Dai for helpful conversations. 
\paragraph{Funding:}
This research is supported by the U.S. Army Research Office grant No. W911NF-24-1-0237.
Z.S. also wants to acknowledge the National Natural Science Foundation of China (12174111). K.W. and T.T. acknowledge support from the JSPS KAKENHI (Grant Numbers 20H00354 and 23H02052) and World Premier International Research Center Initiative (WPI), MEXT, Japan.The magneto-optical measurements supported by the US Department of Energy (DE-FG02-07ER46451) were performed at NHMFL, which is supported by the NSF Cooperative Agreement (Nos. DMR-1644779 and DMR-2128556) and the State of Florida.

\paragraph{Authors contributions:}
Q.W., D.W.S., and Z.S. formulated the main concept of the experiments. Q.W., D.V., A.R., B.V, and J.Y. fabricated the TMD samples. K.W. and T.T. provided high quality hBN material. Q.W., D.V., and J.B. designed and performed the PL experiments in Pittsburgh. X.L. and D.S. performed the magnetic field experiments in Florida. I.V.B. performed the theory calculations. N.Y., Q.W., D.V., and D.W.S. designed the electrical gating system. Q.W., D.V., I.V.B., and D.W.S. wrote the paper.  

\paragraph{Competing interests:} The authors declare that they have no competing interests. 
\paragraph{Data and materials availability:} All data needed to evaluate the conclusions in the paper are presented in the paper and/or the Supplementary Materials.

\clearpage

\bibliography{references.bib}
\clearpage
\title{Supplementary Information for:Proof of Light-Induced Electron Pairing in a Bilayer Structure}
\date{\today}
\maketitle
\beginsupplement
\section{Sample Fabrication}
\label{sect.Fab}

The samples were built on 1.5cm x 1.5cm wafers of 300nm SiO2$\slash$Si substrates. A standard optical lithography process was employed using 650nm thick S1805 photoresist. The lithography was written using a Heidelberg MLA100 Direct Write Lithographer and further developed using AZ400K Developer (1:4).  A two-step process was implemented to define local markers and the back gates in one step; and subsequently the contact pads in the second step. This was done to deposit the thinnest atomically flat ( $\delta h_{rms}<1$~nm) gold back gates while still maintaining thick contact pads sufficient for wire bonding. The thin flat back gates increase the yield and quality of the dry transfers required to build the Quaternion heterostructure. The back gates were fabricated using a Plassys E-beam Evaporation system to deposit 3nm Ti$\slash$ 20nm Au for the back gates and 5nm Ti$\slash$ 120nm Au for the contact pads.

The source 2D material flakes were synthesized using standard mechanical exfoliation of bulk hBN and MoSe$_2$ bulk crystals on polydimethylsiloxane (PDMS) using Nitto blue tape ~\cite{castellanos2014deterministic}. The bulk MoSe$_2$ crystals were purchased from ‘HQ Graphene’ and '2D Semiconductors', and the hBN was supplied by the National Institute for Materials Science (Tsukuba, Japan). The PDMS is then carefully aligned and pressed down onto the substrate. The MoSe$_2$ was released from the PDMS at a temperature between 35°C to 40°C. This was done to increase the yield of the dry transfer as well as to produce high quality MoSe$_2$ monolayer transfers. In contrast, hBN does not require heat for high quality transfers. Once the heterostructure stack is complete, e-beam lithography using 350nm thick PMMA is implemented to define electrical contacts between contact pads and the individual MoSe$_2$ layers presented in the bi-layer Quaternion heterostructure stack presented in the paper.

\begin{figure*}[!ht]
\centering
\includegraphics[width=0.7\linewidth]{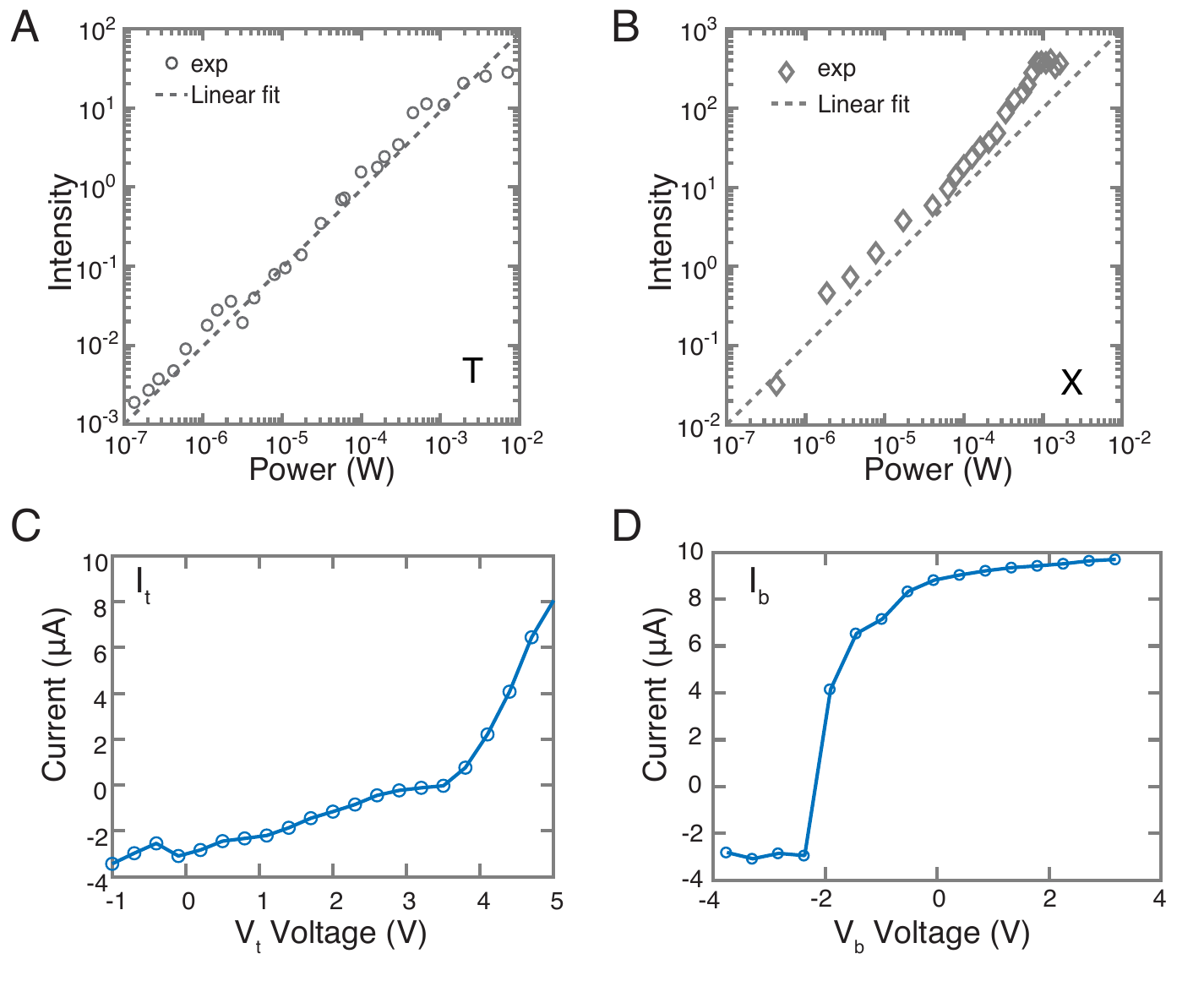}
\caption{\textbf{Additional data for the PL spectra with electrical gating:}  \textbf{(A)} Trion power series at  $V_t = 2$~V and $V_b = -3$~V. \textbf{(B)} The exciton power series at  $V_t = 2$~V and $V_b = -3$~V.  The data of both (A) and (B) were measured on the sample sample as Figures 3C and 3D of the main text. 
  \textbf{(C)} The $I-V$ curve for the first sample, when back gate voltage $V_b=+4$~V, corresponding to the same conditins as the top-gate sweep in Figure~\ref{Fig3}A of the main text. \textbf{(D)} The $I-V$ curve for a the second sample, used for
 Figure~\ref{fig.sample2} below, for $V_t = 0$ and sweeping the back gate voltage, under the same conditions as the data for Figure~\ref{fig.sample2}.}
\label{Fig_S2}
\end{figure*}

\begin{figure*}[!ht]
\centering
\includegraphics[width=0.7\linewidth]{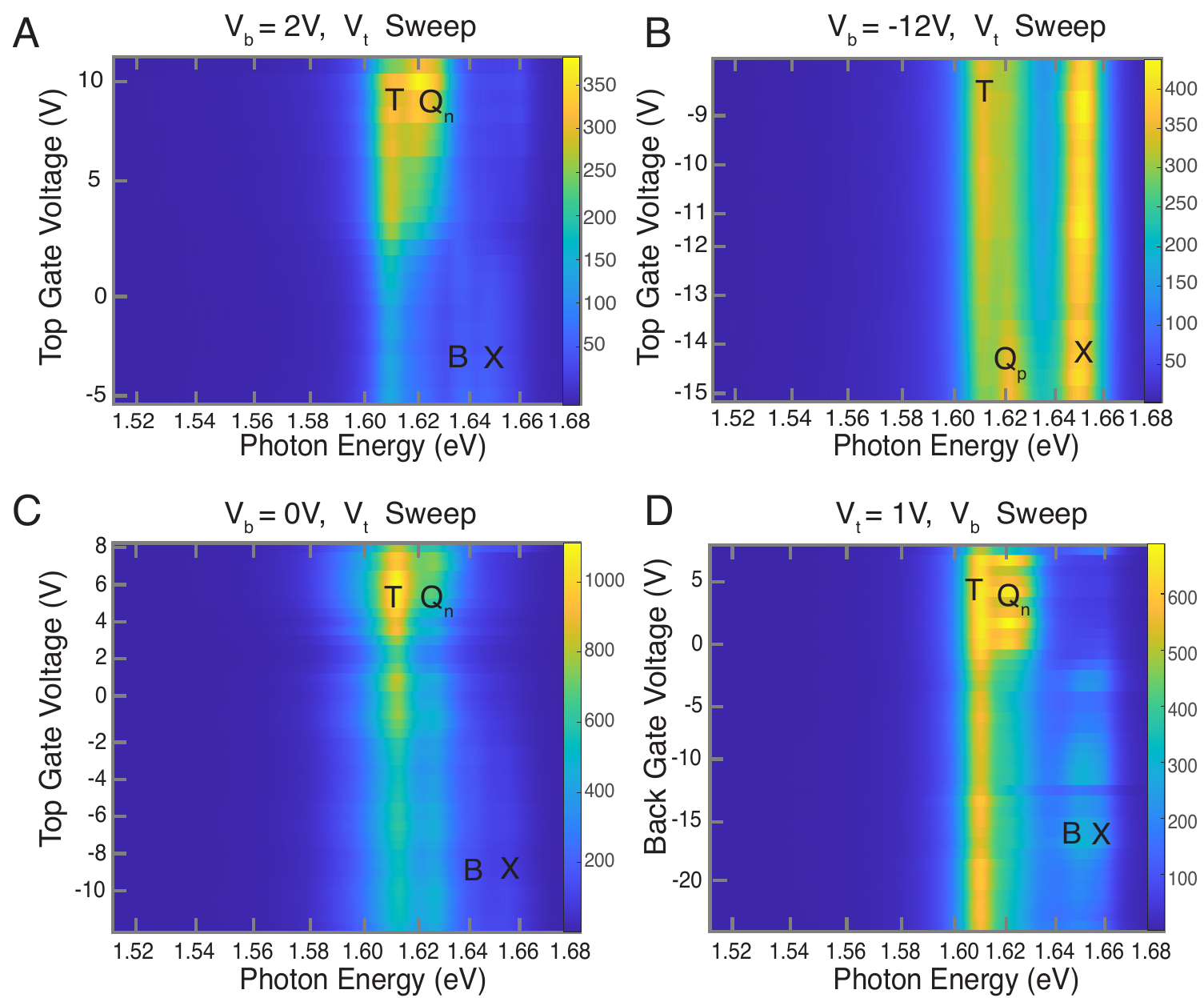}
\caption{{\textbf{Additional variable doping measurements:}  \textbf{(A)} The PL measured under constant back gate doping ($V_b=2$~V) and changing $V_t$ from -5 V to 10 V.  \textbf{(B)} The PL measured under constant back gate doping ($V_b=-12$~V) and changing $V_t$ from -15 V to -9 V. The positive quaternion ($Q_p$) is observed under these conditions.  \textbf{(C)}  The PL measured under constant back gate doping ($V_b=0$) and changing $V_t$ from -10 V to 8 V. The quaternion intensity is very weak in this case. \textbf{(D)}  The PL measured under constant top gate doping ($V_t=1V$) and changing $V_b$ from -20 V to 5~V.}}
\label{figextra}
\end{figure*}

\begin{figure*}[!htbp]
\centering
\includegraphics[width=0.7\linewidth]{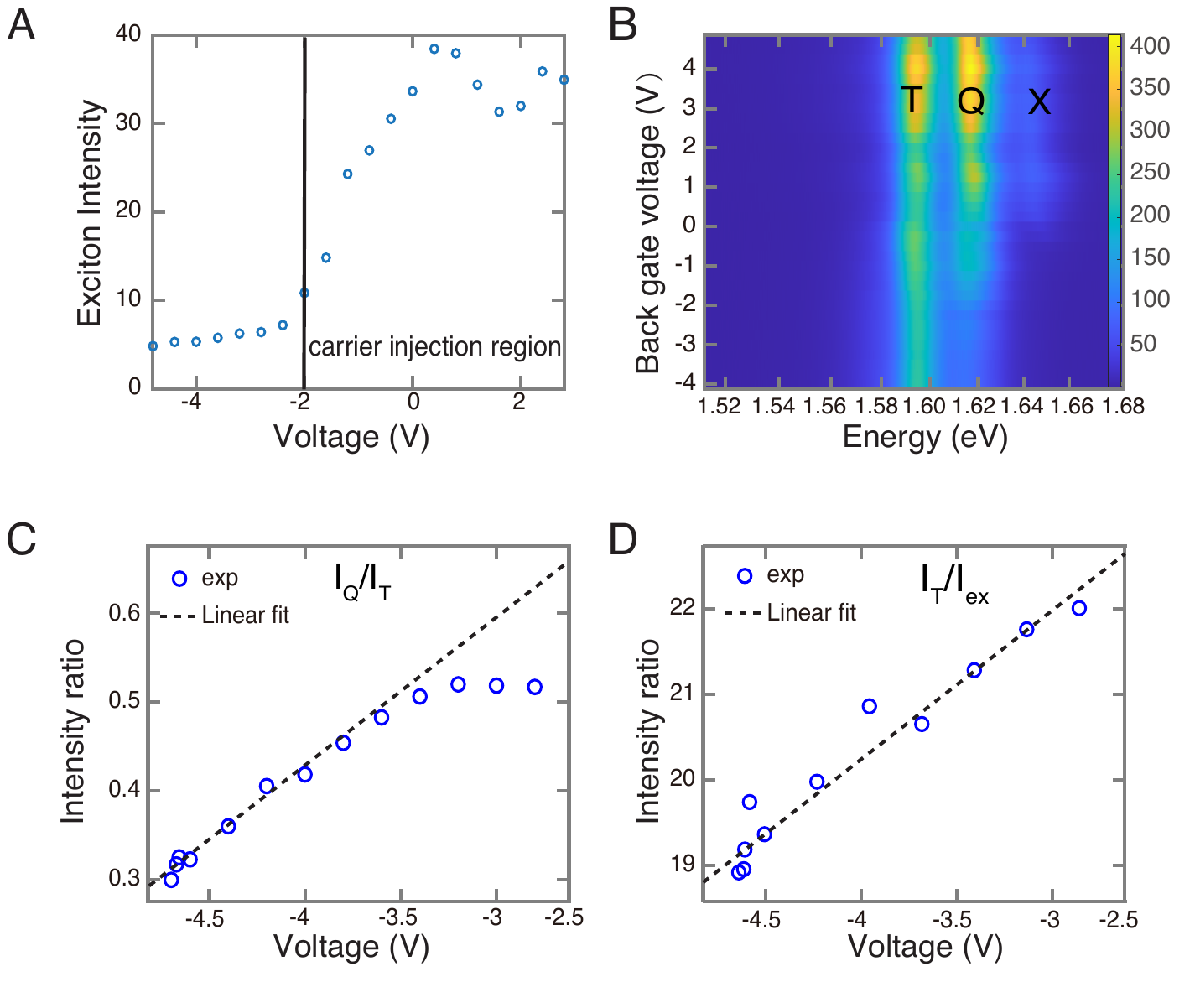}
\caption{\textbf{Additional PL data from a second gated quaternion sample:} \textbf{(A)}  The exciton intensity as a function of the back gate, for the top gate grounded.  As seen in this figure, the
exciton intensity dramatically jumps up above $V_b \simeq -2$~V, indicating that exciton creation by electron/hole injection is occurring. \textbf{(B)} The PL spectrum as a function of the back gate voltage for the grounded top gate.  \textbf{(C)} The ratio of the quaternion to trion intensity in the range where the exciton intensity is nearly constant. \textbf{(D)} The ratio of the trion to exciton intensity over the same voltage range.}
\label{fig.sample2}
\end{figure*}

\section{Additional Voltage-Dependent PL Data}

As discussed in the main text, the middle layer of the structure is defined as the ground in all experiments. We therefore have two voltages to sweep, the back gate and top gate. Figure~\ref{fig.sample2} shows several additional sweeps of the voltage.

An overall trend can be seen, that when the top and back gates are both negative relative to ground, this will create a trap for the electrons in the grounded middle layer, enhancing negative trion formation, and by extension, negative quaternion formation. Similarly, if both are positive, this will trap holes in the middle layer, enhancing the creation of positive trions and quaternions.  

Figures~\ref{Fig_S2}A and \ref{Fig_S2}B give additional pump-power dependence data for the sample discussed in the main text, under the same experimental conditions. Figure~\ref{Fig_S2}A shows the exciton intensity, and Figure~\ref{Fig_S2}B shows the trion intensity, at  $V_t = 2V$ and $V_b = -3V$. As seen in this figure, both are linear with pump power, as expected. Figure \ref{Fig_S2}C gives a typical $I-V$ curve for this sample, corresponding to the same conditions as Figure 3A of the main text, and Figure~\ref{figextra} shows additional voltage sweeps for this sample.

Figure~\ref{Fig_S2}D and Figure~\ref{fig.sample2} show data from a second structure similar to the structure used for the data of Figures 2 and 3 of the main text. This structure had a poor connection to the middle layer, leading to a floating gate that wasn't fully grounded. The effect of this was that exciton generation via current injection occurred at large voltage differences between the top and back gates. This is seen in Figure~\ref{fig.sample2}(A), where the exciton emission intensity jumps up as voltage is increased, even while the optical pump power remains the same. This behavior corresponds to the large jump of current in Figure~\ref{Fig_S2}D. This means that in this voltage range, we do not expect a standard mass-action equation to apply, because there will be a highly nonequilibrium population dynamic.  In Figures~\ref{fig.sample2}C and \ref{fig.sample2}D, we plot the ratio of the quaternion to trion intensities, and the ratio of the trion to exciton intensities, in just the range where the exciton intensity stays nearly constant. As seen in these figures, the same behavior occurs as shown for the first sample, shown in Figure~3 of the main text.

\section{Group Theoretical Analysis of the Spin Structure}\label{sect.spin}

The eigenstates and selection rules of 2D materials have often been analyzed in terms of pseudospin, but it is instructive here to analyze the states in terms of the full group theory.  As discussed in Ref.~\cite{xiao2012coupled}, the full symmetry group of TMD monolayers in Koster notation is D$_{3h}$. At the K-valleys which correspond to the lowest energy gap, the symmetry is reduced to C$_{3h}$. We will stick to the case of negative trions and quaternions, which corresponds to the intrinsically $n$-doped sample used in the magnetic field measurements.

The conduction band states arise from the $z^2$ $d$-orbital, which is $\Gamma_1$ in $D_{3h}$, while the valence band states arise from the $xz$ and $yz$ $d$-orbital states, which have $\Gamma_5$ symmetry in $D_{3h}$, and the $(x\pm iy)^2$ $d$-orbital states, which correspond to $\Gamma_6$. These latter are the topmost valence bands. 

When electron spin, which corresponds to the $\Gamma_7$ in $D_{3h}$, is included, the conduction band states become $\Gamma_7$, while the topmost valence band states split into $\Gamma_7 \otimes \Gamma_6 = \Gamma_{8} \oplus \Gamma_9$. On reduction to $C_{3h}$ symmetry in the $K$-valleys, the $\Gamma_{8}$ states become $\Gamma_{9} \oplus \Gamma_{10}$ (spin $m_J = \pm \frac{5}{2}$), while the $\Gamma_{9}$ states become $\Gamma_{11} \oplus \Gamma_{12}$ (spin $m_J = \pm \frac{3}{2}$). The spin-orbit energy gives a value of around 100 meV between the $\frac{3}{2}$ and $\frac{5}{2}$ valence states. The ``A'' bright excitons are the lowest state, formed of the $\pm \frac{3}{2}$ holes and the $\pm \frac{1}{2}$ electrons, corresponding to
\begin{eqnarray}
\begin{array}{ll}
|-\frac{1}{2}\rangle |+\frac{3}{2}\rangle, & \hspace{1cm} m_J = +1 \hspace{.5cm} (K~\mbox{valley})\\
\\
|+\frac{1}{2}\rangle |-\frac{3}{2}\rangle, & \hspace{1cm} m_J = -1 \hspace{.5cm} (K'~\mbox{valley}).
\end{array}
\end{eqnarray}
There are also two ``dark'' A-excitons corresponding to 
\begin{eqnarray}
\begin{array}{ll}
|+\frac{1}{2}\rangle |+\frac{3}{2}\rangle & \hspace{1cm} m_J = +2\\
\\
|-\frac{1}{2}\rangle |-\frac{3}{2}\rangle
& \hspace{1cm} m_J = -2.
\end{array}
\end{eqnarray}
which cannot emit single photons.

The intralayer trion-bound states are formed by adding another electron in the A-exciton state: 
\begin{eqnarray}
\begin{array}{ll}
|-\frac{1}{2}\rangle |+\frac{3}{2}\rangle |-\frac{1}{2}\rangle, & \hspace{0.5cm} m_J = +\frac{1}{2} \hspace{.5cm} (K~\mbox{valley})\\
\\
|+\frac{1}{2}\rangle |-\frac{3}{2}\rangle |+\frac{1}{2}\rangle, & \hspace{0.5cm} m_J = -\frac{1}{2} \hspace{.5cm} (K'~\mbox{valley}).
\end{array}
\end{eqnarray}
Each of these emits a photon with the same handedness as an A-exciton in the same valley. 

Quaternions are formed of a trion in one layer and a free electron in the adjacent valley. The spin product of the $J=\frac{1}{2}$ trions and the $S = \frac{1}{2}$ extra electron gives a spin-0 singlet and a spin-1 triplet. As discussed in the main text, the triplet state is the lowest state, corresponding to the electron in the separated well having the opposite spin of the two electrons in the trion. These three states are written
\begin{eqnarray}
\begin{array}{ll}
\left(|-\frac{1}{2}\rangle |+\frac{3}{2}\rangle |-\frac{1}{2}\rangle\right)|+\frac{1}{2}\rangle, & \hspace{1cm} m_J = +1  \\
\\
\frac{1}{\sqrt{2}}\left(|-\frac{1}{2}\rangle |+\frac{3}{2}\rangle |-\frac{1}{2}\rangle\right)|-\frac{1}{2}\rangle
 & \hspace{1cm} m_J = 0\\
+ \frac{1}{\sqrt{2}}\left(|+\frac{1}{2}\rangle |-\frac{3}{2}\rangle |+\frac{1}{2}\rangle\right)|+\frac{1}{2}\rangle,
\\
\\
\left(|+\frac{1}{2}\rangle |-\frac{3}{2}\rangle |+\frac{1}{2}\rangle\right)|-\frac{1}{2}\rangle, & \hspace{1cm} m_J = -1 .
\end{array} 
\label{quattrip}
\end{eqnarray}
As seen in this analysis, two of the states emit with 100\% circular polarization in one direction or the other, while the $m_J$ state is a 50\% superposition of both handednesses.  As with the trions, the energy of the emitted photon depends on the energy of the two remaining electrons. These have opposite spin for the two $m_J = \pm 1$ quaternions, and a superposition of aligned spins in with $m_J=\pm 1$ for the $m_J = 0$ quaternion. 

In addition to the spin-triplet quaternion, there is also a spin singlet at higher energy, corresponding to the antisymmetric superposition
\begin{eqnarray}
\begin{array}{ll}
\frac{1}{\sqrt{2}}\left(|-\frac{1}{2}\rangle |+\frac{3}{2}\rangle |-\frac{1}{2}\rangle\right)|-\frac{1}{2}\rangle
- \frac{1}{\sqrt{2}}\left(|+\frac{1}{2}\rangle |-\frac{3}{2}\rangle |+\frac{1}{2}\rangle\right)|+\frac{1}{2}\rangle
, & \hspace{1cm} m_J = 0 
\end{array}
\label{quatsing}
\end{eqnarray}
This will also be 50\% allowed to emit either handedness of polarization.

We can also examine the spin structure of the indirect transitions, that is, the interlayer transitions. The $m_J = \pm 1$ quaternion states in (\ref{quattrip}) are forbidden to emit single photons corresponding to a hole in one layer and electron in the other layer, but the $m_j =0$ state can emit such a photon.  There is evidence for this in the very weak line seen at low energy, labeled as $Q'$, in Fig.~\ref{FigS5}). There will also be two allowed emission lines with $J = \pm 1$ corresponding to a single exciton made of a hole in one layer and an electron in the other layer. Finally, an interlayer trion is also possible, consisting of an exciton in one layer and a free electron in the other layer. Indirect emission is possible from two such trion states: 
\begin{eqnarray}
\begin{array}{ll}
\left(|-\frac{1}{2}\rangle |+\frac{3}{2}\rangle \right)|-\frac{1}{2}\rangle, & \hspace{1cm} m_J = +\frac{1}{2}  \\
\\
\left(|+\frac{1}{2}\rangle |-\frac{3}{2}\rangle \right)|+\frac{1}{2}\rangle
, & \hspace{1cm} m_J = -\frac{1}{2}.  
\end{array} 
\end{eqnarray}
Both the interlayer exciton and interlayer trion states have emission that shifts with $B$-field similarly to the intralayer versions. Only the quaternion $m_J = 0$ state does not have this linear shift.

Finally, we look at the spin structure of a biexciton made of two A-excitons in the same layer. The ground state has two excitons of opposite spin \cite{ostatnicky2005spin}, which we write as
\begin{eqnarray}
\begin{array}{ll}
\left(|+\frac{1}{2}\rangle |-\frac{3}{2}\rangle \right)\left(|-\frac{1}{2}\rangle |+\frac{3}{2}\rangle \right), & \hspace{1cm} m_J = 0.
\end{array}
\end{eqnarray}
Either of the two excitons can emit a photon. When a photon is emitted, the final state is an exciton of the opposite handedness. The energy of the photon is therefore the difference between the unshifted biexciton and the exciton energy shifted by magnetic field, which gives an energy shift of the emitted photon.

\section{Theoretical Model for the Quaternion Energy Shifts}

 To elucidate the behavior of quaternion states observed in magnetostatic field experiments, we constructed a theoretical framework based on the spin-Hamiltonian formalism.  This approach was originally pioneered by Landau, Gor’kov and Pitaevski, Holstein and Herring, in their studies of molecular binding and magnetism \cite{herring1962critique,landau2013quantum}. This approach provides a foundational understanding of a broad spectrum of physical phenomena, including nuclear magnetic resonance (NMR) and electron spin resonance (ESR) effects 
 ~\cite{slichter2013principles}. Its foundation was laid back in the 1950's by titans such as Anderson~\cite{anderson1950antiferromagnetism,anderson1955considerations}, Feynman~\cite{FeynmanQM} and Herring~\cite{herring1962critique,herring1964asymptotic}. They applied this method to study the hyperfine splitting of the ground state of atomic hydrogen, hydrogen molecular ion and hydrogen molecule as well as collective spin phenomena in solids~\cite{feynman2018statistical}. A complicated total Hamiltonian is not necessary to analyze the behavior of a complex molecular type entity in a constant magnetic field. Utilizing the stepwise averaging of all the operators of coordinates and momenta in it, apart from the spin operators involved, it can be reduced to an effective spin Hamiltonian to grasp the essence of the Zeeman effect on the system~\cite{landau2013quantum}.

We consider our quaternion complex being built of the spin-1/2 intralayer trion in a bottom monolayer coupled to a spin-1/2-like charge carrier in the top monolayer, consistent with both spin-structure analysis in the previous section and earlier works~\cite{bondarev2018complexes,bondarev2021crystal,zhou2021bilayer}. Considering the negative quaternion for definiteness, we have a negatively charged four-particle complex closely resembling the hydrogen, muonium or positronium atoms, with the only difference that its nucleus (trion) is negatively charged while its stability is supported by image charges in the metallic substrate. The spin-Hamiltonian for this negative quaternion system can be obtained from the spin-Hamiltonian of the hydrogen atom in a constant magnetic field, reviewed in full detail in Ref.~\cite{FeynmanQM}, by merely reversing the sign of the proton magneton and the sign of the positive hyperfine splitting constant there, which is proportional to the product of the electron and proton charges. By relabeling the terms accordingly, one then obtains
\begin{equation}
\hat{H}= -A\,\bm{\sigma}_e\!\cdot\bm{\sigma}_T+\mu_e\textbf{B}\!\cdot\!\bm{\sigma}_e+\mu_T\textbf{B}\!\cdot\!\bm{\sigma}_T\,.
\label{main}
\end{equation}
Here, $A$ and $\mu_{e,T}\!=\!(g_{e,T}/m_{e,T})\,\mu_B$ are the positive constants to represent the electron-trion hyperfine (spin-spin) coupling and their respective magnetic moments with their effective masses (in units of the free electron mass $m_0$) and $g$-factors included, $\mu_B\!=\!|e|\hbar/(2m_0c)\!=\!0.0579~\mbox{meV/T}$ is the Bohr magneton, and $\bm{\sigma}_{e,T}\!=\!(\sigma_x,\sigma_y,\sigma_z)_{e,T}$ are the Pauli matrices of the electron and trion spin-1/2 subspaces,
\begin{equation}
\sigma_{x_{e,T}}=\left(\begin{array}{cc}0 & 1 \\ 1 & 0 \end{array}\right),\;\;\;\sigma_{y_{e,T}}=\left(\begin{array}{cc}0 & -i \\ i & 0 \end{array}\right),\;\;\;\sigma_{z_{e,T}}=\left(\begin{array}{cc}1 & 0 \\ 0 & -1 \end{array}\right).
\label{Pauli}
\end{equation}

\begin{figure}
\begin{center}
\includegraphics[width=.95\linewidth]{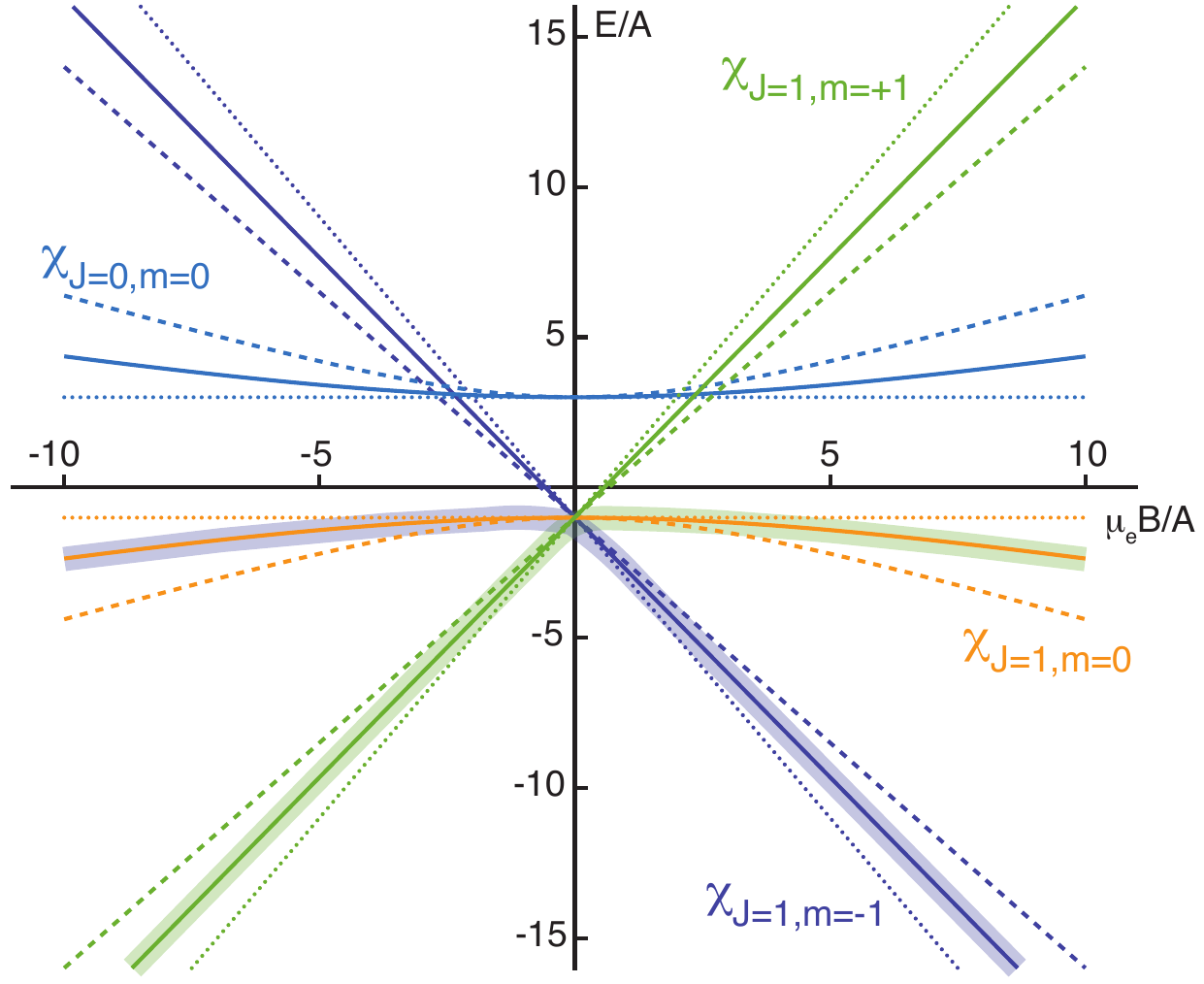}
\end{center}
\caption{\textbf{Theoretical prediction of the magnetic field dependence of the quaternion spin states}: Triplet (orange, green, purple) and singlet (blue) quaternion energy levels with their respective eigenvectors marked in a constant magnetic field of both orientations as given by Eqs.~(\ref{mu})-(\ref{E34}) for $\zeta=0.5, 0.73~\mbox{and}~1$ shown by dashed, solid and dotted lines, respectively. The green and purple bands indicate the lowest energy quaternion states that emit photons of right and left circular polarizations respectively as discussed in the previous section.}
\label{figS4}
\end{figure}

With the $z$-quantization axis directed along the magnetic field $\textbf{B}$ and the general spin-state wave function
\begin{gather}
|\chi\rangle = C_1|+\rangle_{e}|+\rangle_{T}+C_2|+\rangle_{e}|-\rangle_{T}\\ 
+C_3|-\rangle_{e}|+\rangle_{T}+C_4|-\rangle_{e}|-\rangle_{T},\;\;\; \nonumber\\
|+\rangle=\left(\!\!\begin{array}{c}1\\0\end{array}
\!\!\right),
\hspace{1cm} \;\;|-\rangle=\left(\!\!\begin{array}{c}0\\1\end{array}\!\!\right),
\label{spinstate}
\end{gather}
where the $C_i$ for $i= 1$ through 4 are unknown coefficients, the matrix of the spin-Hamiltonian (\ref{main}) takes the following form
\begin{equation}
\begin{array}{ccccc}
\hskip1.57cm~\vline & |+\rangle_{e}|+\rangle_{T} & |+\rangle_{e}|-\rangle_{T} & |-\rangle_{e}|+\rangle_{T} & |-\rangle_{e}|-\rangle_{T}\\[0.25cm]
\hline
|+\rangle_{e}|+\rangle_{T}\;\;\;\vline & -A+\mu & 0 & 0 & 0 \\[0.25cm]
|+\rangle_{e}|-\rangle_{T}\;\;\;\vline & 0 & A+\mu^\prime & -2A & 0 \\[0.25cm]
|-\rangle_{e}|+\rangle_{T}\;\;\;\vline & 0 & -2A & A-\mu^\prime & 0 \\[0.25cm]
|-\rangle_{e}|-\rangle_{T}\;\;\;\vline & 0 & 0 & 0 & -A-\mu
\label{Hmatrix}
\end{array}
\end{equation}
with
\begin{gather}
\mu=(1+\zeta)\,\mu_eB,~~~\mu^\prime=(1-\zeta)\,\mu_eB,~~~ \notag \\
\zeta=\frac{\mu_T}{\mu_e}=\frac{g_{T}}{g_{e}}\frac{m_{e}}{m_{T}}=\frac{g_{T}}{g_{e}}\frac{m_{e}}{m_{h}+2m_{e}},
\label{mu}
\end{gather}
where the electron $g$-factor and electron-hole effective masses can be obtained from the literature, $g_e\approx2.5$~\cite{oreszczuk2023enhancement}, $m_e\approx0.88$, $m_h\approx0.74$~\cite{liu2021exciton}, yielding $g_T\approx5.18$ from Zeeman shift reported~\cite{li2014valley}, to give an estimate $\zeta\approx0.73$, accordingly. On diagonalization we get the following set of eigenvalue problem solutions (eigenvalues and eigenvectors):
\begin{equation}
\begin{array}{ll}
E_{1,2}= -A\pm\mu,\;\;\;
& |\chi_{_{1,2}}\rangle=|\pm\rangle_{e}|\pm\rangle_{T}=|\chi_{_{11,1-1}}\rangle, \\
E_{3,4}=A\Big(1\mp2\sqrt{1+x^2}\Big),\;\;\;
&|\chi_{_{3,4}}\rangle=C_{\pm}|\chi_{_{10}}\rangle\mp C_{\mp}|\chi_{_{00}}\rangle,\;\;\; 
\end{array}
\end{equation}
with
\begin{equation}
x=\frac{|\mu^\prime|}{2A}=\,\frac{|1-\zeta|}{2}\frac{\mu_eB}{A},
\label{E34}
\end{equation}
where
\begin{equation}
|\chi_{_{10,\:00}}\rangle=\frac{1}{\sqrt{2}}\Big(|-\rangle_{e}|+\rangle_{T}\pm|+\rangle_{e}|-\rangle_{T}\Big) 
\label{singlettriplet}
\end{equation}
and
\begin{equation}
C_{\pm}=\sqrt{\frac{1}{2}\Big(1\pm\frac{1}{\sqrt{1+x^2}}\Big)},\;\;\;|C_{+}|^2+|C_{-}|^2=1.
\label{Cpm}
\end{equation}
Here, for convenience, the normalized spin eigenfunctions are also written in the form $|\chi_{_{Sm_s}}\rangle$ to indicate the actual spin configurations. The triplet with $S=1,m_s=\pm1,0$ is the ground state and the singlet with $S=0,\,m_s=0$ is the excited state, as shown in Fig.~\ref{figS4}. The compound electron-trion system is assumed to be in the $s$-state of its relative orbital motion, whereby the quantum number $S$ herein is equal to the total angular moment $J$ of the previous section.

\section{Additional Magneto-optical Measurements}

\begin{figure*}
\centering
\includegraphics[width=0.7\linewidth]{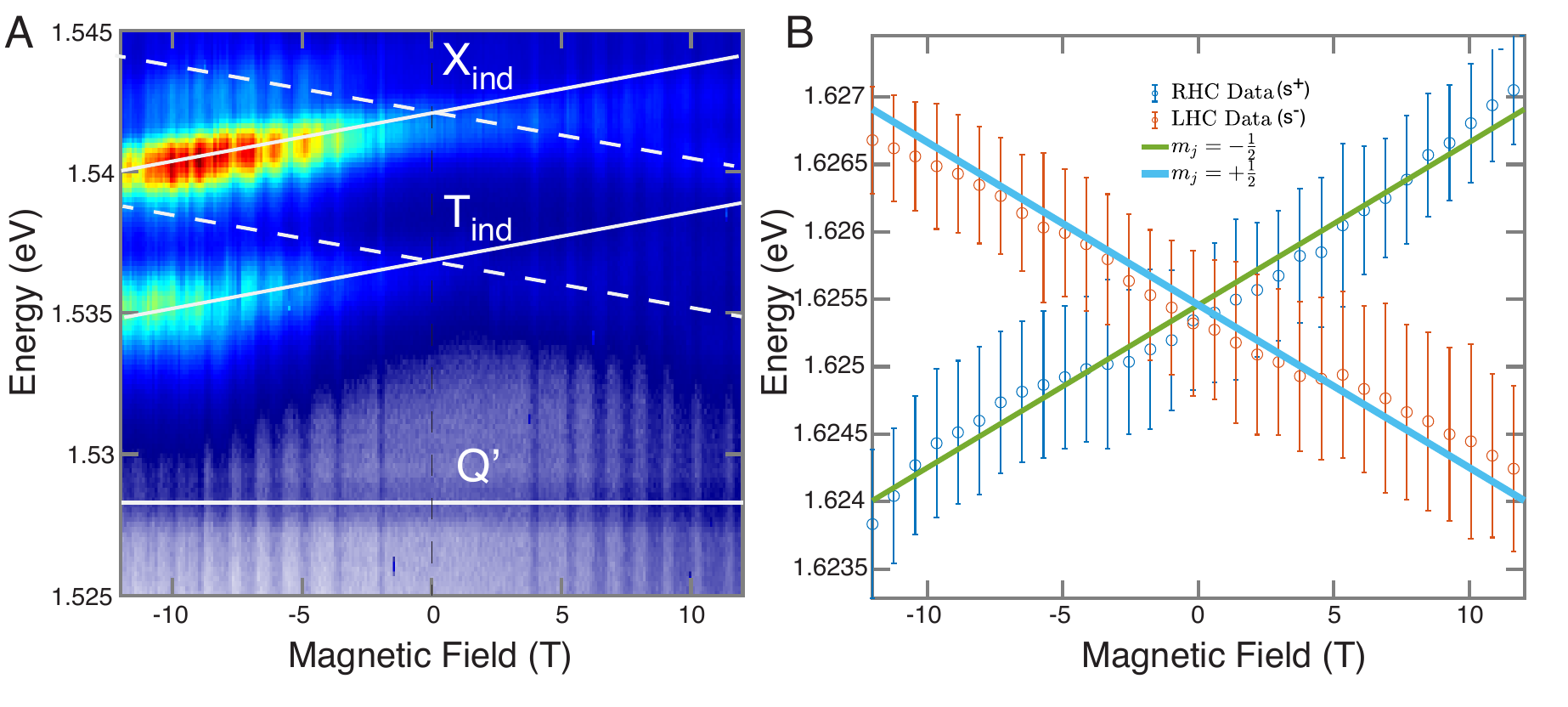}
\caption{\textbf{Magnetic response of other excitonic states}. \textbf{(A)} Right-circular polarized PL spectrum as a function of magnetic field  at longer wavelength for the same conditions as Figure 4A of the main text. We identify these lines as $X_{\rm ind} =$ the spatially indirect exciton, $T_{\rm ind}=$ the spatially indirect trion, and $Q'=$ the indirect transition from the same $m_J=0$ quaternion species as the one that gives the intralayer (direct) emission in Figure 4A of the main text.  The white solid lines give linear shifts of the lines, and the dashed white lines show the corresponding shift for the opposite handedness. \textbf{(B)}  The trion energy shifts with the magnetic field, measured on the same sample as Figure 4. The RHC and LHC data symmetrically cross near the zero field. The Zeeman shift extracted from the fit can be found in the Table \ref{tab:2}. }
\label{FigS5}
\end{figure*}

In addition to the four bright states seen in Figure~\ref{Fig4}, we also see states present at lower photon emission energy. Figure~\ref{FigS5} shows the emission spectrum as a function of magnetic field for right-circularly polarized light with the same conditions as Fig.~\ref{Fig4} in the main text but at much lower photon energies. This range of the spectrum is well known \cite{jiang2021interlayer,horng_observation_2018} to give the emission of interlayer species; the energies are lower because the intrinsic doping typically gives a band offset of the band gaps of the two layers in the bilayer structure. 

As discussed in the group theory section above, there are three allowed transitions for spatially indirect PL emission, all of which require tunneling through the barrier between the layers. Two of these are for distinct bound states that are not the same as those giving intralayer emission, namely the indirect exciton and indirect trion. These are seen in the strong circularly polarized emission at higher energies in Figure~\ref{FigS5}, with the predicted linear Zeeman shifts; each of these has two spin states. In addition, there is expected to be a second emission line from the $m_J =0$ quaternion. This emission is suppressed not only by the tunneling through the barrier that is needed but also by the selection rule of the spatial orbitals; although the transition is an allowed spin transition, spatial wave functions in the two different layers have different parity. The line with nearly zero slope, labeled $Q'$ in Figure~\ref{FigS5}, is very weak but can be tentatively identified as this emission.

The PL measurements shown in Figure~\ref{Fig4} and Figure~\ref{FigS5} were analyzed using a least-squares fit to a sum of Lorenztian peaks to extract the contributions from each of the excitonic complexes. The centers of these Lorentzians were then fitted using the theoretical models described above to extrapolate the Zeeman shifts observed in Figure~\ref{FigS5} and Figure~\ref{Fig4}. For the quaternion, we have three curves to fit with four parameters due to the thermal population suppression of the $|\chi_{J=0\;;\;m_J=0}\rangle$ singlet state. This means that the hyperfine splitting energy parameter $A$ cannot be fit independently of the ground state energy $E_0$ and the combined Zeeman shift ($\mu' = |1-\zeta|\,\mu_B$). From our best fit of the quaternion state, the value of $A$ can vary $\pm$0.2 meV without significant ($<5\%$) change in the chi-squared value. 

\begin{table}[htb!]
    \centering
    \caption{\textbf{Values of the parameters of the quaternion model used to fit the PL Data}}
    \begin{tabular}{c c}
         &  \\
         \hline
         Fitting Parameter & Value\\
         \hline
         Ground State Energy $E_0$ (eV)& 1.612\;$\pm$\;0.001\\
         Hyperfine Splitting Energy A (meV)& 0.346\;$\pm$\;0.2\\
         $\mu=(1+\zeta)\mu_e$ (meV/T)&0.235\;$\pm$\;0.05\\
         $\mu'=(1-\zeta)\mu_e$ (meV/T)&0.078\;$\pm$\;0.058\\
         $\zeta=\mu_T/\mu_B$&0.78\;$\pm$\;0.21\\
         \hline
         & 
    \end{tabular}
    \label{tab:1}
\end{table}

The values of the binding energy and Zeeman shifts for the various exciton complexes are shown in Table~\ref{tab:2}. For all of the magnetic field fits, we used an offset of $B_0 = -0.49$ T, which is consistent with a small intrinsic magnetic polarization in our sample, which was doped with rhenium, a magnetic impurity known to give weak magnetization \cite{kochat2017re}.  Se vacancies in these monolayer materials can also introduce weak ferromagnetism \cite{ma2011electronic,yang2023emerging}.
\begin{table*}[htb!]
    \centering
    \caption{\textbf{Fitted energies for the exciton, biexciton, trion, and indirect exciton lines versus those measured from literature}}
    \begin{tabular}{c c c c c}
         &  \\
         \hline
		Excitonic & Exciton Energy & Zeeman Shift & Lit. Exciton Energy & Lit. Zeeman Shift\\
        Complex&(eV)&(meV/T)&(eV)&(meV/T)\\
        \hline
        Exciton&1.654&0.204\;$\pm$\;0.02&1.655\cite{li2014valley}&0.12\;$\pm$\;0.01\cite{li2014valley}\\
        Biexciton&1.643&0.165\;$\pm$\;0.02&N/A&N/A\\
        Trion&1.625&0.111\;$\pm$\;0.01&1.625\cite{li2014valley}&0.12\;$\pm$\;0.01\cite{li2014valley}\\
        Ind. Exciton&1.541&0.122\;$\pm$\;0.01&1.550\cite{horng_observation_2018}&N/A\\
		\hline       
         & 
    \end{tabular}
    \label{tab:2}
\end{table*}
\begin{figure*}[!htp]
\centering
\includegraphics[width=0.7\linewidth]{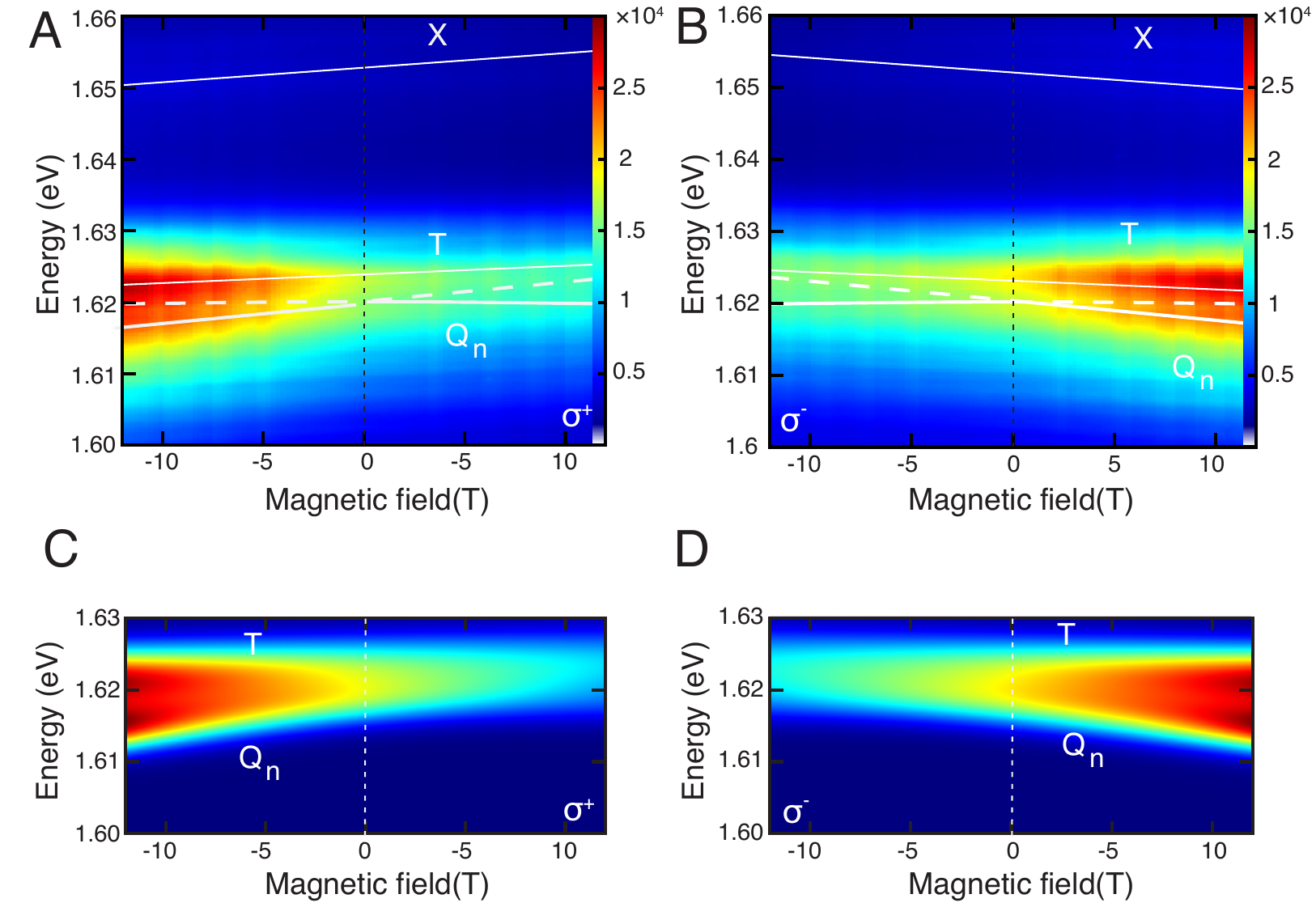}
\caption{\textbf{Additional magneto-PL measurements:} \textbf{(A)} PL spectrum as a function of magnetic field for a right-circular polarized emission. (The vertical striations correspond to laser intensity fluctuations.) X: exciton line; T: trion line;  Q: quaternion line. The white lines indicate linear shifts of the lines with magnetic field. \textbf{(B)} PL spectrum for left-circular polarized PL for the same conditions, with the slopes of the line shifts, but time-reversed $B \rightarrow -B$. \textbf{(C)} Simulated data for the right-circular emission from the quaternion triplet, using the linear energy shifts shown in (A) above, weighted by the Maxwellian occupation factor $e^{-E/k_BT}$, with $T = 35$~K, and line width of 4.6 meV.
\textbf{(D)} Simulated data for the left-circular emission from the quaternion triplet, using the linear energy shifts shown in (B) above, weighted by the Maxwellian occupation factor $e^{-E/k_BT}$, with $T = 50$~K. The elevated bath temperature reflects the consumption of helium during sequential measurements. }
\label{FigS6}
\end{figure*}

A second quaternion sample was also examined under a magnetic field, exhibiting a significantly smaller energy gap between the quaternion and trion states. The gap was sufficiently narrow that we cannot extract the quaternion state from the spectrum. To address this, we employed the same simulation algorithm used for Figure 4C to model a scenario where the trion and quaternion states are closely spaced within the spectral domain. The energy separation between the quaternion and trion states is governed by the thickness of the hexagonal boron nitride (hBN) layers. This sample featured a notably thinner middle hBN layer and a considerably thicker bottom layer, which contributed to the observed spectral characteristics.

An additional point of note is the sequence of measurements: left-hand circularly polarized (LHC, $\sigma^-$) emission was recorded following right-hand circularly polarized (RHC, $\sigma^+$) emission. The sequential nature of these measurements led to a slight increase in the sample temperature during the LHC data collection due to the liquid helium running low during the Florida lab measurements.

\end{document}